# Single-shot optical imaging through scattering medium using digital in-line holography

Vinu R. V[1], Kyoohyun Kim[2], Atul S. Somkuwar[1], YongKeun Park[2], and Rakesh Kumar Singh[1]

[1]Applied and Adaptive Optics Laboratory, Department of Physics, Indian Institute of Space Science and Technology (IIST), Valiamala, Trivanrdum, 695547, Kerala , India

[2]Department of Physics, Korea Advanced Institute of Science and Technology, Daejeon, 305-701, Republic of Korea

Correspondence should be addressed to RKS (krakeshsingh@iist.ac.in)

**Abstract:**

**Non-invasive and single-shot holographic imaging through complex media is technically challenging due to random light scattering which significantly scrambles optical information. Recently, several methods have been presented to address this issue. However, they require complicated measurements of optical transmission matrices, or existing techniques do only retrieve intensity information. Here we propose and experimentally demonstrate a holographic approach for single-shot imaging through a scattering layer based on digital in-line holography in combination with the autocorrelation of the speckle intensity. Using a simple optical configuration and experimental procedure, the proposed method enables to retrieve the complex amplitude image of an object located at arbitrary planes behind scattering media. The technique has potential applications in biomedical imaging, deep tissue microscopy, and 3D imaging through turbid media.**

## INTRODUCTION

Imaging through complex media has attained significant attention because this is a fundamentally important physical problem and also have diverse applications in focusing, imaging, biophotonics, and nanotechnology. Light propagating through complex media suffers significant multiple scattering due to inhomogeneous distributions of refractive index in scattering media, which makes it difficult to deliver optical information using conventional imaging techniques. The complex-valued object information is scrambled in the scattered grainy intensity pattern called the laser speckle. Techniques for extracting information from laser speckles are a tedious process, and it starts with the invention of laser. Initial work on coherent propagation of light through scattering media started from the work of Leith and Upatneik in 1966, where they used a hologram to recover the object information through a scattering medium[1]. A significant number of techniques for static and dynamic imaging through random media based on holographic principles were developed in later years[2–7]. Recent developments in optical coherence tomography arise as a powerful method of imaging which uses only the ballistic light and separates the diffused light[8]. Use of wave front shaping technique to suppress the turbidity and thereby to enhance the tomography images was also reported[9]. Recent advances in the use of scattering medium to focus, shape and compress waves by controlling the many degrees of freedom of incident waves gives a new realm in imaging and focusing in turbid media[10,11]. Availability of high sensitive and fast performing megapixel digital sensors and modulators provide more flexibility in imaging and wave front shaping. Wave front shaping techniques by the use of spatial light modulator and by exploiting the turbidity of the scattering layer, researchers were able to reconstruct the image through opaque materials and improve the focusing and spatial resolution beyond the diffraction limit[12–16]. Demonstrations of techniques based on optical phase conjugation and progress in controlling the spatial and temporal components of optical waves offer greater flexibility in suppressing the turbidity and focus light through turbid media [17–21]. Most of the recent advances in imaging through a scattering medium are invasive and time-consuming since they require a wave front shaping mechanism behind the scattering medium or a detector or a long acquisition sequence and also sensitive to misalignment in certain cases.

Very recently remarkable achievements towards the non-invasive imaging through scattering layers and imaging of moving objects through dynamic scattering media were demonstrated by applying correlation approach[22-25]. The technique presented by Bertolotti et. al uses an angular speckle correlation to image a fluorescent object inside an opaque layer[22]. The angular scanning of speckle pattern results in a long acquisition time of images and limits the object to be stationary for the complete acquisition period. The non-invasive single-shot imaging technique presented by Katz et. al is capable of overcoming the limitation of long acquisition period by capturing the single-shot speckle pattern[23]. Imaging through dynamic turbid media using shower curtain effect and speckle correlography is demonstrated very recently[24]. More recent demonstration for moving objects hidden by thick scattering medium provide access to the spatially dependent dielectric constant of the moving objects[25]. For retrieval of the object information, these techniques employ phase retrieval algorithms with real-valued object constraint. Moreover, the depth information of the object behind the random scattering medium is not available. Progress in holographic techniques in recent years based on numerical reconstruction capability attains a new level of interest due to its huge potential in three-dimensional and complex field imaging. Particularly role of in-line holography is widely appreciated due to its experimental simplicity and technique has been applied in studying the tissue optical properties, cell type characterization, dynamic structure of cell membranes and the angle resolved light scattering of micron-sized objects[26-30]. Even though these techniques have a unique position in biological applications, it is not able to give a non-invasive imaging through the scattered medium where the complete information of the object is scrambled into a speckle pattern. Techniques for microscopic imaging and quantitative phase contrast imaging of objects in turbid media by digital holography is demonstrated by exploiting the Doppler frequency shift experienced by the photons scattered by the flowing colloidal particles[31].

To circumvent depth imaging issue behind the random scattering medium, we recently developed an off-axis holography, i.e. recorded hologram behind the random scattering medium, which reconstruct the complex-valued object behind the scatterer. However, recording the off-axis hologram behind the scatterer is not always feasible[32,33]. In this communication, we propose and experimentally demonstrate a non-invasive single-shot imaging technique to recover the complex-valued objects lying behind the scattering medium using the principles of in-line (Gabor type) holography in combination with the speckle autocorrelation. The use of in-line holography for the imaging of the object lying behind the random scattering medium has the potential to image complex-valued objects and provides depth information of the sample behind the scattering layer. Even though the scatterer scrambles the object information and generates the complex speckle pattern, our technique utilizes the randomness of the speckle pattern to recover the in-line hologram formed behind the scattering medium. The recovery of the digital in-line hologram from the speckle autocorrelation gives the opportunity for numerical reconstruction of the object information at the desired plane behind the scattering medium. To demonstrate the potential of the proposed technique, imaging of various objects at different depths behind the scattering layer were illustrated. Principle and experimental implementation of the proposed technique is explained below.

**MATERIALS AND METHODS**

**Principle:**

Let us consider a complex-valued object located at a distance, $-z = d$, behind a random medium. When the object is illuminated by a coherent beam, the scattered light from the object propagates through the scattering medium. If the scattered beam from the object is small in compared to the un-scattered one, the holographic diffraction is dominant and it results into the formation of an in-line (or Gabor type) hologram[27,28]. The wave amplitude reaching the scattering medium located at a distance $d$ from the object is represented as

$$U(\hat{x}, \hat{y}) = U_r(\hat{x}, \hat{y}) + U_s(\hat{x}, \hat{y}) \qquad , \quad (1)$$

where $(\hat{x}, \hat{y})$ is the spatial coordinate, $U_r(\hat{x}, \hat{y})$ is the un-scattered wave and $U_s(\hat{x}, \hat{y})$ is the scattered wave. The corresponding intensity at the scattering medium is given by

$$I(\hat{x}, \hat{y}) = |U_r(\hat{x}, \hat{y})|^2 + |U_s(\hat{x}, \hat{y})|^2 + U_r^*(\hat{x}, \hat{y})U_s(\hat{x}, \hat{y}) + U_r(\hat{x}, \hat{y})U_s^*(\hat{x}, \hat{y}) . \qquad (2)$$

The first two terms in the right-hand side of Eq. (2) represent the intensity of the un-scattered and scattered beam, respectively. Third and fourth terms represent the intensity terms of real and virtual images in the in-line hologram. The intensity term in Eq. (2) contains the in-line hologram and illuminates an inhomogeneous complex scattering layer. In conventional holography, object information is reconstructed from the intensity pattern by applying different numerical reconstruction techniques[27]. However, such reconstruction techniques cannot be directly applied in this case due to scrambling of the hologram information by the scattering medium.

A schematic representation of the formation of in-line hologram and its obstruction by the scattering layer is shown in Fig.1. This inhomogeneous scattering layer present in the path of the in-line hologram obstructs the coherent beam due to the random scattering and scrambles the object information. A direct intensity recording of the scattered light gives a speckle pattern at the recording plane and object information is completely scrambled. In our proposed experimental strategy, we retrieve the digital in-line hologram at the scattering plane from a single-shot speckle pattern recorded at the far field from the scattering medium and this gives us the opportunity to measure the complex field behind the medium, and thereby retrieving the object information at the desired plane using back propagation techniques. The far field plane is located at the Fourier plane of a lens of a focal length $f$ as shown in Fig 1. The autocorrelation of the co-variance of the speckle intensity is connected to the modulus square of the far field diffraction pattern of illumination function $I(\hat{x}, \hat{y})$ [34], and is given by

$$\langle \tilde{I}(x,y)\tilde{I}(x+\gamma, y+\delta)\rangle \propto |F(\gamma, \delta)|^2, \tag{3}$$

where $F(\gamma, \delta) = \iint I(\hat{x}, \hat{y}) \exp\left[-j\frac{2\pi}{\lambda f}(\hat{x}\gamma + \hat{y}\delta)\right] d\hat{x}d\hat{y}$, the far field diffraction of illumination function, $\langle ... \rangle$ represents the ensemble average, $\tilde{I}(x, y) = I(x, y) - \langle I(x, y)\rangle$ is a variation of intensity around the mean over the plane of observation. $I(\hat{x}, \hat{y})$ is the illumination function at the scattering medium and $(x, y)$ and $(x + \gamma, y + \delta)$ are the spatial coordinates in the observation plane. The autocorrelation function of the speckle intensity provides the modulus square of complex field $F(\gamma, \delta)$ as given in Eq. (3) and phase part is lost. Losing the phase information makes it impossible to recover the illumination function[35]. Modulation of the Fourier spectrum information in association with phase retrieval algorithms can be used for image synthesis without depth information for certain class of objects[22,23,36]. However in this work, we apply a new approach to recovering the lost phase of Fourier spectrum from the speckle interferometry approach rather than phase retrieval. To circumvent phase recovery challenge, we followed an interferometry approach. An independent reference speckle pattern is generated from a known illumination function and allows the coherent superimposition of the reference speckle field coming from the in-line hologram's arm. The intensity distribution of the resultant speckle field at the observation plane is given by

$$I(x, y) = |E_O(x, y) + E_R(x, y)|^2 \tag{4}$$

where $E_O(x, y)$ is the speckle field contribution of the object through the scattering medium and $E_R(x, y)$ is the reference speckle field. The far field diffraction contribution arises from the illumination of an off-axis reference beam at an independent scattering medium is given by

$$F_R(\gamma, \delta) = \iint circ\left(\frac{\hat{x} + \hat{y} - (\hat{x}_g + \hat{y}_g)}{a}\right) \exp\left[-j\frac{2\pi}{\lambda f}(\hat{x}\gamma + \hat{y}\delta)\right] d\hat{x}d\hat{y}, \tag{5}$$

where $\hat{x}_g$ and $\hat{y}_g$ represents the lateral shift of the reference beam of radius '$a$' on the reference scattering medium.

Since the scatterers used to realize the in-line hologram and reference speckles are different, we are justified in taking the contribution from the mixed term zero, i.e. $\langle E_O(x, y)E_R^*(x+\gamma, y+\delta \approx 0)\rangle$. Therefore, the autocorrelation of this resultant speckle pattern at the observation plane results into an interferogram given by

$$|F(\gamma, \delta)|^2 = |F_O(\gamma, \delta)|^2 + |F_R(\gamma, \delta)|^2 + F_O(\gamma, \delta)F_R^*(\gamma, \delta) + F_O^*(\gamma, \delta)F_R(\gamma, \delta), \tag{6}$$

where $F_O(\gamma, \delta)$ is the far field diffraction pattern arises from the illumination function $I(\hat{x}, \hat{y})$ at the scattering medium and $F_R(\gamma, \delta)$ is the far field diffraction pattern arises from the reference speckle pattern. We would like to emphasize and highlight here difference in Eq. (2) and Eq. (5). Eq. (1) is an in-line hologram but Eq. (6) is an off-axis (Leith-Upatniek type) hologram from the autocorrelation of the co-variance of the speckle intensity. Using a Fourier transform method of fringe analysis technique[32], the complex field $F_O(\gamma, \delta)$ of the object field at the Fourier plane can be separated out from other unwanted terms. The complete retrieval of complex diffraction pattern $F_O(\gamma, \delta)$ at the observation plane provides us the potential to recover the in-line hologram at the scattering medium. In an experimental implementation, ensemble average is replaced by the spatial averaging for the spatially stationary random field[37]. The requirement of spatial stationarity over the observation plane is fulfilled by the Fourier transforming lens.

The in-line hologram at the scattering medium is retrieved from the complex far field diffraction pattern through a Fourier relation and is given as:

$$I(\hat{x}, \hat{y}) = \iint F_S(\gamma, \delta) \exp\left[ j\frac{2\pi}{\lambda f}(\hat{x}\gamma + \hat{y}\delta) \right] d\gamma d\delta, \qquad (7)$$

where $I(\hat{x}, \hat{y})$ is the intensity distribution corresponding to the illumination function which contains the digital in-line hologram information of the complex-valued object located at distance d from the scattering medium. From Eq. (7) it is clear that the digital in-line hologram which illuminates the random inhomogeneous media can be recovered from the complex far field diffraction pattern from single shot recording of speckle pattern.

The digital in-line hologram recovered at the random media gives the potential to retrieve the 3-D reconstruction of complex-valued object information using numerical techniques. The reconstruction quality of the images from in-line hologram is degraded by the twin image formation as explained as the main drawback of in-line holography[28]. In order to eliminate the virtual image or the out of focus image formed in the in-line holography reconstruction, the recovered digital hologram was numerically propagated to a conjugate plane $z = d$ from the recovered hologram plane by Rayleigh-Sommerfield back propagation[30]. By proper identification of the edge of the out of focus image, the inside region of the image is replaced with the average value of the background or outside the edge. The resulting field is then back propagated to the desired plane ($z = -d$) where the real object or in focus object is present. Thus, the complex field retrieved allows the complete reconstruction i.e. both amplitude and phase information of the object at the desired plane thereby providing a depth resolved imaging of the object hidden behind the random inhomogeneous medium.

**Experimental Implementation:**

Experimental implementation of the proposed method is demonstrated in Fig.2. A linearly polarized He-Ne laser beam with the wavelength of 632.8nm (Melles Griot, 25-LHP-925) is spatially filtered using a microscope objective (L1) and a pin hole (A1) combination. The filtered beam is collimated using a lens (L2) of focal length 150 mm and is using as the input beam to a Mach-Zehnder type interferometer. The collimated beam is divided into two by a beam splitter (BS1). The reflected beam from BS1 is folded by mirror M1 and illuminates the samples hidden behind the scattering layer. Note that, we have carried out experimental verification of the proposed technique for two types of samples namely transmissive and reflective (Supplementary). USAF 1951 resolution chart and polystyrene beads are used as transmitting type objects for experimental demonstrations. The USAF 1951 resolution chart is from Edmund Optics, USA. The polystyrene beads are immersed in Olympus immersion oil and sandwiched between two cover slips. Reflective type objects were introduced in the experimental setup using a reflective type spatial light modulator (SLM) by replacing mirror M1 with beam splitter BS3, polarizer P and SLM (Supplementary information Fig. 1).The SLM used in the experimental implementation is LC-R 720 (reflective) from Holoeye. It is based on LCOS microdisplay with a resolution of 1280×768 pixels with a pixel pitch of 20μm.The object that to be imaged is displayed in the SLM and the reflected beam from SLM carries the object information.

The reflected beam (carries reflecting type object) or transmitted beam (carries transmitting type object) from the object suffers a holographic diffraction and forms an in-line hologram and scattered from the random diffuser medium (Ground glass diffuser- GG1). The ground glass diffuser used is DG20-120-MD - Ø2" (Thorlabs) with 120 grit value. The light get scattered from the scattering medium carries the holographic information but is completely scrambled in the speckle pattern. The transmitted beam from BS1 acting as the reference arm of the Mach-Zehnder interferometer is used to generate the known independent reference speckle pattern. This beam is folded by a mirror M2 then illuminates an independent ground glass GG2 through a microscope objective MO. The beam from MO is pointed to fall in GG2 in an off-axis position in order to provide a linear phase term into the autocorrelation of the reference speckle pattern as explained by Eq. (5). The speckle patterns from the two arms of the interferometer are coherently superimposed using a non-polarizing beam splitter BS2. The combined field is Fourier transformed using a lens L3 of focal length 75mm. The resultant speckle field at the Fourier transform plane is captured by a high resolution CCD monochrome camera (Prosilica GX 2750). The camera is of 14 bit with 2750×2200 pixels and a pixel pitch of 4.54μm. The high resolution and large total pixel size of the CCD camera allows us to implement spatial averaging technique by replacing the ensemble averaging, thereby provides the potential to develop a single–shot imaging technique.

**RESULTS AND DISCUSSION**

To explore the validity of our proposed technique, we implemented the experimental setup. Let us first turn attention to a reflective type object which is displayed onto an SLM in association with polarizer P (Supplementary). The SLM is illuminated by a He-Ne laser source and the random speckle pattern is recorded at the Fourier transform plane by a CCD. For experimental demonstration an object '5' of dimension $0.94\,mm \times 0.54\,mm$ is inserted in the experimental setup using the SLM at a distance of 288mm behind the ground glass diffuser (GG1). The holographic diffraction of the object on propagation results the formation of an in-line hologram. The size of the in-line hologram is controlled by an aperture of size 5 mm in order to control the size of speckles at the camera. The size of the aperture is chosen in such a way that the speckle grain size at the recording plane is larger than two CCD pixels in order to meet the sampling criterion. Figure 3(a) shows the speckle intensity at the CCD plane, which exhibit random patterns in nature and has no visible relation to the object displayed in the SLM. Autocorrelation of the co-variance of the speckle intensity pattern results into interference fringes at the CCD plane as described in Eq. (6) that contains the complex field function, $F_O(\gamma,\delta)$. Using Fourier transform method, $F_O(\gamma,\delta)$ at the CCD plane is separated out from the other unwanted terms (Supplementary information, Fig. 2).The intensity $I(\hat{x}, \hat{y})$ of the digital in-line hologram at the scattering medium can be recovered from the complex field function, $F_O(\gamma, \delta)$, by making use of the Fourier result as described in Eq. (7). Thus the in-line hologram at the scattering plane is efficiently recovered from the random speckle pattern and is shown in Fig. 3(b). By using Rayleigh-Sommerfield back propagation and twin image removal as explained in the above section, the complex-valued object information at the desired plane from the scattering medium can be reconstructed from the recovered in-line hologram. Figures 3(c) and 3(d) show the reconstructed amplitude and phase distribution of the object '5' at a distance of 288mm behind the scattering medium.

For the quantitative analysis of the reconstruction capability of the proposed technique, we evaluated the parameters such as visibility and reconstruction efficiency for two cases with and without the scattering medium. Visibility ($v$) is defined as the ratio of average intensity level of signal region to the average intensity level of background region in the reconstructed image. The signal region is estimated by using a global threshold approach. The reconstruction efficiency ($\eta$) gives the ratio of the measured power in the signal region of the reconstructed image to the sum of the measured power in signal and background region. The visibility and reconstruction efficiency parameters for the reconstructed image of object '5' with scattering medium are 15.6218 and 0.9398, respectively. In order to verify the reconstruction capability with scattering medium, we estimated the parameters without scattering medium i.e. the direct recording of the in-line hologram at a distance of 288mm from the SLM by the CCD camera (Supplementary information, Fig.3). The corresponding reconstruction parameters $v$ and $\eta$ are 16.2050 and 0.9419, respectively. A good agreement between imaging through scattering layer and a direct in-line hologram imaging manifests the potential of the proposed technique. We also tested this technique for another reflective object 'V' and moreover imaging of this letter is performed at different z-planes. In order to check the performance the object 'V' of dimension $\Box$ $0.8mm \times 0.8mm$ is displayed at various planes behind the scattering medium. The complex amplitude recovered at various planes behind the scattering medium from 100 mm to 600 mm with an interval of 100 mm is shown in Fig. 4. The reconstruction parameters for various planes behind the scattering medium for the cases with and without scattering medium are estimated and are shown in Table.1. It is clear from Fig. 4 and Table 1 that the quality of reconstruction is deteriorated for short and long distances between the object and the scatterer. When an object is too far from the scatterer, high spatial frequency components are not able to reach the limited size scatterer window and thus to the detector which leads to the deterioration of the reconstructed image quality. Also in short distances, the diffraction pattern of the object reaching the scattering medium is very narrow and thus it is difficult to fulfill the requirement of delta function correlation at the scattering medium and leads to deterioration of the image quality. This feature is very much similar to the reconstruction of in-line hologram in the free space.

Furthermore potential of the proposed technique is tested for transmission case. For this purpose, we use a transmission-type object as our samples. A USAF 1951 resolution chart is lying 50 mm behind the scattering medium GG1. A coherent beam illuminates the object and forms an in-line hologram at the scattering medium. The scattered object beam is coherently superposed with the reference speckle from GG2 and the resultant speckle field is recorded at the CCD plane. As in the previous case the in-line hologram at the scattering medium is recovered from the speckle field using autocorrelation of resultant speckle intensity by invoking Fourier transform operation and filtering process to the interferogram as explained in Eq. (6). Experimental results are shown in Fig. 5 for two different cases where the beam illuminates group 0 and group 1 area of the USAF resolution chart. The resultant speckle field recorded at the CCD plane is shown in Fig. 5(a) and 5(b) for group 0 and group 1 area of USAF resolution chart respectively. The corresponding reconstructed amplitude distributions at a distance 50 mm behind the scattering medium are shown in Fig. 5 (c) and 5(e) and the phase distributions are shown in Fig. 5(d) and 5(f) respectively. In order to measure the reconstruction quality of the obtained results, we estimated the reconstruction parameters from the recovered image at actual depth behind the scattering medium. The visibility of the retrieved image for group 0 and group 1 area of USAF resolution chart are found to be 18.9183 and 14.7490, respectively and the corresponding reconstruction efficiency are 0.9498 and 0.9365, respectively. A good extent of recovery of both amplitude and phase distribution at actual depth behind the scattering medium shows the potential of the technique in depth resolved imaging of real world objects through scattering medium.

Moreover to check the potential of the technique in imaging of micron scale objects behind the scattering medium and to test the strength in biological application we used micron-sized polystyrene beads as our sample objects. The polystyrene beads are immersed in Olympus immersion oil and sandwiched between two cover slips. The sample is placed at a distance of 70 mm behind the scattering layer. The polystyrene beads are randomly distributed between the cover slips and it forms an in-line hologram at the scattering medium. This in-line hologram illuminates the scatterer and generates the random speckle pattern. This speckle pattern is superposed with reference speckle pattern and recorded at the CCD plane. As mentioned in the above case, the in-line hologram at the scattering medium is recovered from the random speckle pattern using autocorrelation of the resultant speckle intensity by applying Fourier fringe approach. Recorded speckle intensity pattern at the CCD plane is shown in Fig. 6(a). The amplitude and phase distribution obtained for a set of polystyrene beads which are randomly distributed between two cover slips are shown in Fig. 6(b) and 6(c) respectively. An efficient depth resolved recovery of both amplitude and phase distribution of polystyrene beads explains the possibility of implementing the technique in imaging through turbid media and biomedical related applications.

## CONCLUSIONS

In conclusion, we experimentally demonstrated a non-invasive single-shot imaging technique through scattering layers by recovering the digital in-line hologram at the scattering medium. The technique utilizes the randomness of the speckle pattern and making use of speckle autocorrelation to recover phase of the Fourier spectrum and consequently recover the in-line hologram from speckle. The recovery of digital in-line hologram from the random speckle pattern gives the opportunity of numerical reconstruction of complex-valued object information at the actual depth. Since the technique recovers the in-line hologram the technique has potential in three-dimensional imaging of objects through scattering layers also. Experimental description of the recovery of amplitude and phase of different reflective and transmissive type objects were demonstrated and a quantitative analysis of reconstruction is performed by analyzing parameters such as visibility and reconstruction efficiency. Thus our technique has a remarkable achievement in true non-invasive and depth resolved imaging through complex media, thereby have robust practical applications in biomedical imaging, imaging through turbid media etc.

**REFERENCES**

1 Leith,Emmett N; Upatnieks J. Holographic imagery through diffusing media. *J Opt Soc Am* 1966; **56**: 523.

2 Goodman JW, Huntley WH, Jackson DW, Lehmann M. Wavefront reconstruction imaging through random media. *Appl Phys Lett* 1966; **8**: 311–313.

3 Kogelnik H, Pennington K. Holographic imaging through a random medium. *J Opt Soc Am* 1968; **58**: 273–274.

4 Hillman TR, Yamauchi T., Choi W., Dasari RR, Feld MS, Park Y., Yaqoob Z. Digital optical phase conjugation for delivering two-dimensional through turbid media. *Sci. Rep.* 2013; 3: 1909.

5 Singh AK, Naik DN, Pedrini G, Takeda M, Osten W. Looking through a diffuser and around an opaque surface: a holographic approach. *Opt Express* 2014; **22**: 7694–7701.

6 Chen Y, Chen H, Dilworth D, Leith E, Lopez J, Shih M *et al.* Evaluation of holographic methods for imaging through biological tissue. *Appl Opt* 1993; **32**: 4330–4336.

7 Li S, Zhong J. Dynamic imaging through turbid media based on digital holography. *J Opt Soc Am A* 2014; **31**: 480–486.

8 Huang D, Swanson EA, Lin CP, Schuman JS, Stinson WG, Chang W *et al.* Optical coherence tomography. *Science* 1991; **254**: 1178–1181.

9 Yu H, Jang J, Lim J, Park J-H, Jang W, Kim J-Y *et al.* Depth-enhanced 2-D optical coherence tomography using complex wavefront shaping. *Opt Express* 2014; **22**: 7514–7523.

10 Mosk AP, Lagendijk A, Lerosey G, Fink M. Controlling waves in space and time for imaging and focusing in complex media. *Nat Photonics* 2012; **6**: 283–292.

11 Yu H, Park J, Lee K, Yoon J, Kim K, Lee S *et al.* Recent advances in wavefront shaping techniques for biomedical applications. *Curr Appl Phys* 2015; **15**: 632–641.

12 Vellekoop IM, Mosk a P. Focusing coherent light through opaque strongly scattering media. *Opt Lett* 2007; **32**: 2309–2311.

13 Vellekoop IM, Mosk AP. Universal Optimal Transmission of Light Through Disordered Materials. *Phys Rev Lett* 2008; **101**: 120601.

14 Popoff S, Lerosey G, Fink M, Boccara AC, Gigan S. Image transmission through an opaque material. *Nat Commun* 2010; **1**: 1–5.

15 Vellekoop IM, Aegerter CM. Scattered light fluorescence microscopy: imaging through turbid layers. *Opt Lett* 2010; **35**: 1245–1247.

16 Choi Y, Yang TD, Fang-Yen C, Kang P, Lee KJ, Dasari RR *et al.* Overcoming the diffraction limit using multiple light scattering in a highly disordered medium. *Phys Rev Lett* 2011; **107**: 1–4.

17 Yaqoob Z, Psaltis D, Feld MS, Yang C. Optical phase conjugation for turbidity suppression in biological samples. *Nat Photonics* 2008; **2**: 110–115.

18 Lee AJ,Lee J, Park J-H., Park J-H, Park YK. One-wave optical phase conjugation mirror by actively coupling arbitrary light fields in a single0mode reflector. *Phys. Rev. Lett.* 2015; **115**: 153902-253902-5.

19   Aulbach J, Gjonaj B, Johnson PM, Mosk AP, Lagendijk A. Control of light transmission through opaque scattering media in space and time. *Phys Rev Lett* 2011; **106**: 5–8.

20   Katz O, Small E, Bromberg Y, Silberberg Y. Focusing and compression of ultrashort pulses through scattering media. *Nat Phot* 2011; **5**: 372–377.

21   McCabe DJ, Tajalli A, Austin DR, Bondareff P, Walmsley IA, Gigan S *et al.* Spatio-temporal focusing of an ultrafast pulse through a multiply scattering medium. *Nat Commun* 2011; **2**: 447.

22   Bertolotti J, van Putten EG, Blum C, Lagendijk A, Vos WL, Mosk AP. Non-invasive imaging through opaque scattering layers. *Nature* 2012; **491**: 232–234.

23   Katz O, Heidmann P, Fink M, Gigan S. Non-invasive real-time imaging through scattering layers and around corners via speckle correlations. *Nat Photonics* 2014; **8**: 784–790.

24   Newman JA, Luo Q, Webb KJ. Imaging hidden objects with spatial speckle intensity correlations over object position. *Optica* 2016; **3**: 71–74.

25   Edrei E, Scarcelli G. Optical imaging through dynamic turbid media using the Fourier-domain shower-curtain effect. *Phys Rev Lett* 2016; **116**: 1–6.

26   Xu W, Jericho MH, Meinertzhagen I a., Kreuzer HJ. Digital in-line holography for biological applications. *Proc Natl Acad Sci U S A* 2001; **98**: 11301–11305.

27   Molony KM., Hennelly BM., Kelly DP, Naughton TJ. Reconstruction algorithhms applied to in-line Gabor digital holographic microscopy. *Opt. Commun.* 2010; **283**: 903-909.

28   Kreis T. *Handbook of Holographic Interferometry*. Wiley-VCH: Weinheim, 2005.

29   Ding H, Wang Z, Nguyen F, Boppart S A, Popescu G. Fourier transform light scattering of inhomogeneous and dynamic structures. *Phys Rev Lett* 2008; **101**: 1–4.

30   Kim K, Park Y. Fourier transform light scattering angular spectroscopy using digital inline holography. *Opt Lett* 2012; **37**: 4161–4163.

31   Paturzo M., Finizio A., Memmolo P., Puglisi R., Balduzzi D., Galli A., Ferraro P. Microscopy imaging and quantitative phase contrast mapping in turbid microfluidic channels by digital holograpgy. *Lab Chip* 2012; **12**: 3073-3076

32   Singh RK, Vinu RV, Anandraj S. Recovery of complex valued objects from two-point intensity correlation measurement. *Appl Phys Lett* 2014; **104**: 111108.

32   Sing RK, Sharma AM, Das B. Quantitative phase-contrast imaging through a scattering media. *Opt. Lett.* 2014; **39**: 5054-5057.

33   Somkuwar AS., Das B., Vinu RV, Park YK, Singh RK. Non-invasive single-shot 3D imaging through a scattering layer using speckle interferometry. *Optica* (Under review).

34   Goldfischer LI. Autocorrelation function and power spectral density of laser-produced speckle patterns. *J Opt Soc Am* 1965; **55**: 247.

35  Labeyrie A. Attainment of diffraction limited resolution in large telescopes by Fourier analyzing speckle patterns in star image. *Astron. & Astrophys* 1970; 6 : 85-87.

36  Idell PS, Fienup JR, and Goodman RS. Image synthesis from nonimaged laser-speckle pattern: *Opt. Lett.* 1987; **12**: 858-860.

37  Takeda M, Wang W, Naik DN, Singh RK. Spatial Statistical Optics and Spatial Correlation Holography: A Review. *Opt Rev* 2014; **21**: 849–861.

# Table

| Z planes (mm) | With Scattering Medium | | Without Scattering Medium | |
| --- | --- | --- | --- | --- |
| | Visibility(v) | Reconstruction Efficiency($\eta$) | Visibility(v) | Reconstruction Efficiency($\eta$) |
| 100 | 12.7282 | 0.9272 | 16.9175 | 0.9442 |
| 200 | 17.6150 | 0.9463 | 19.7889 | 0.9519 |
| 300 | 17.8059 | 0.9468 | 22.4323 | 0.9573 |
| 400 | 15.8655 | 0.9407 | 16.1322 | 0.9416 |
| 500 | 13.7838 | 0.9324 | 14.6849 | 0.9362 |
| 600 | 12.5867 | 0.9264 | 13.7917 | 0.9323 |

Table 1: Reconstruction parameters for various planes behind the scattering medium for the cases with and without scattering medium

**Figures**

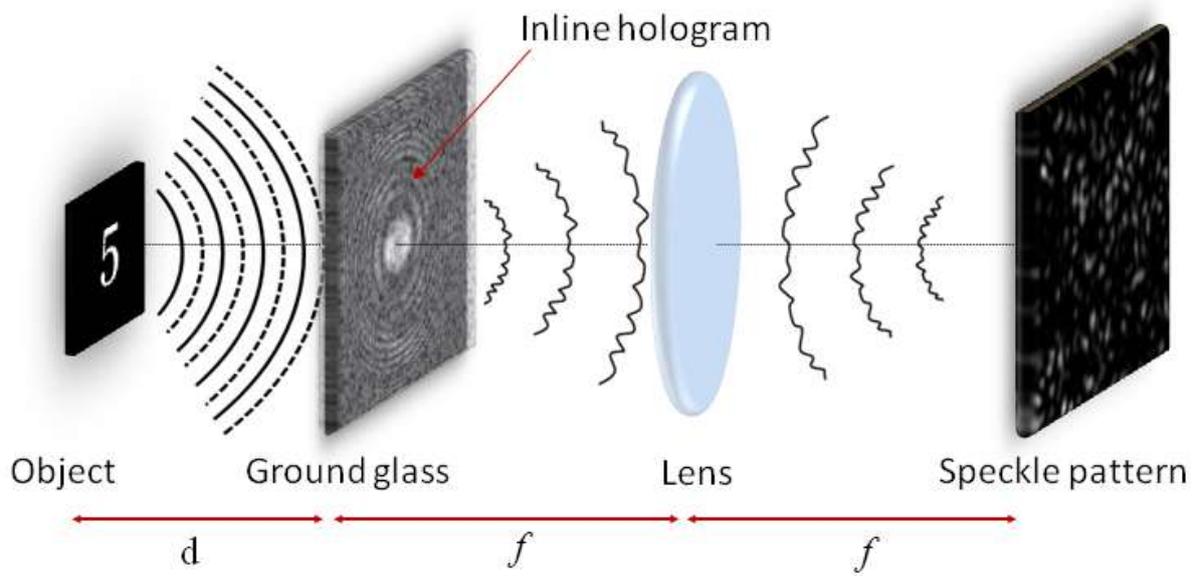

Figure 1: Schematic representation of the formation of inline hologram and its obstruction by scattering medium

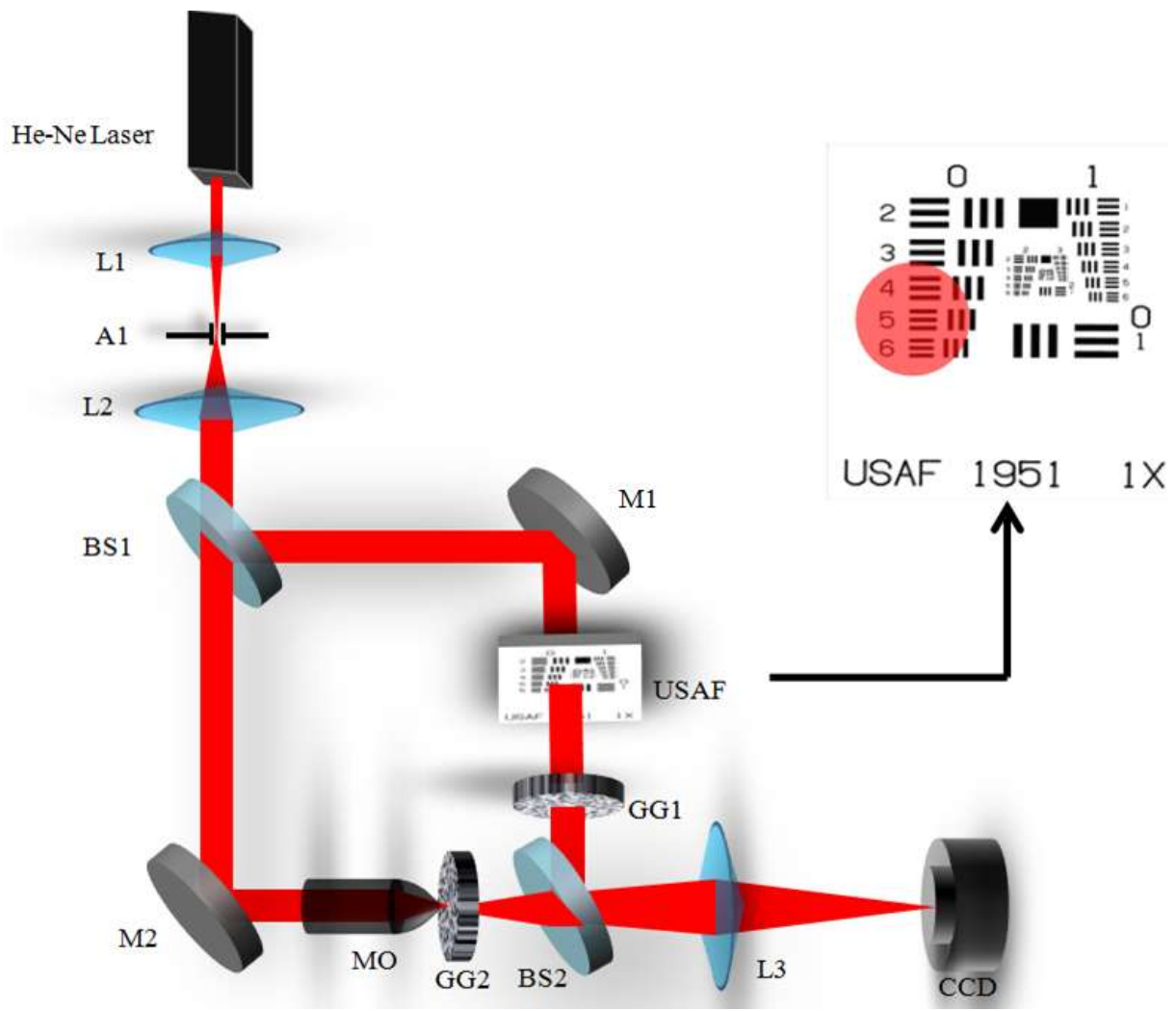

Figure2: Schematic of the proposed experimental setup for the recovery of the object through scattering medium

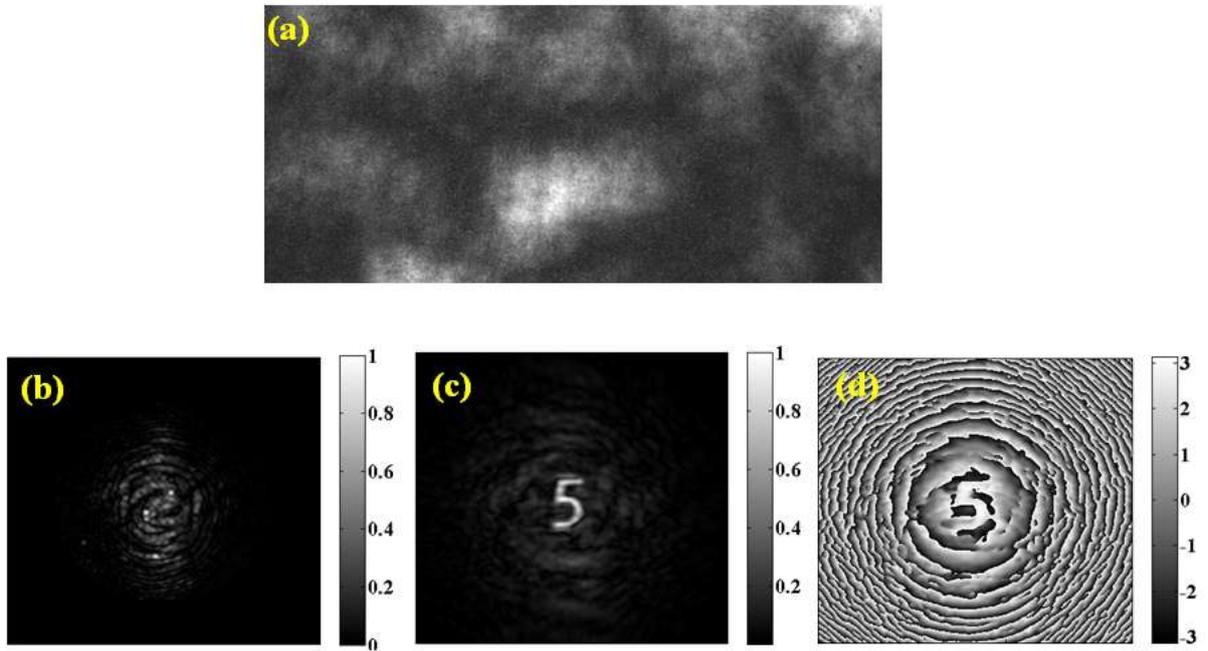

Figure3: Experimental results (a) recorded random speckle pattern at the CCD plane for the object '5' (b) recovered inline hologram at the scattering medium plane (c) recovered amplitude distribution and (d) recovered phase distribution

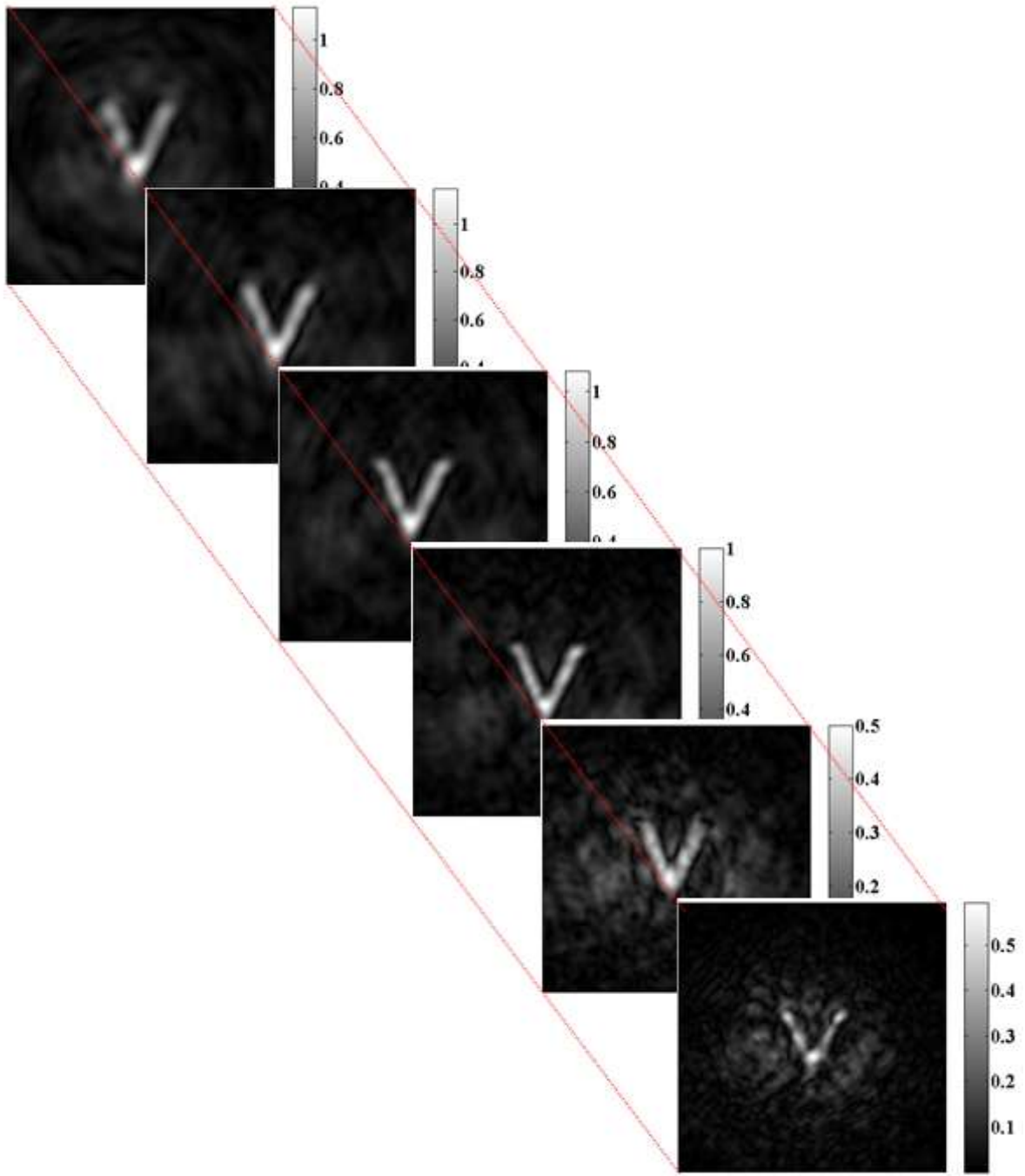

Figure4: Complex amplitude images recovered at various planes behind the scattering medium for an object 'V' displayed in the SLM

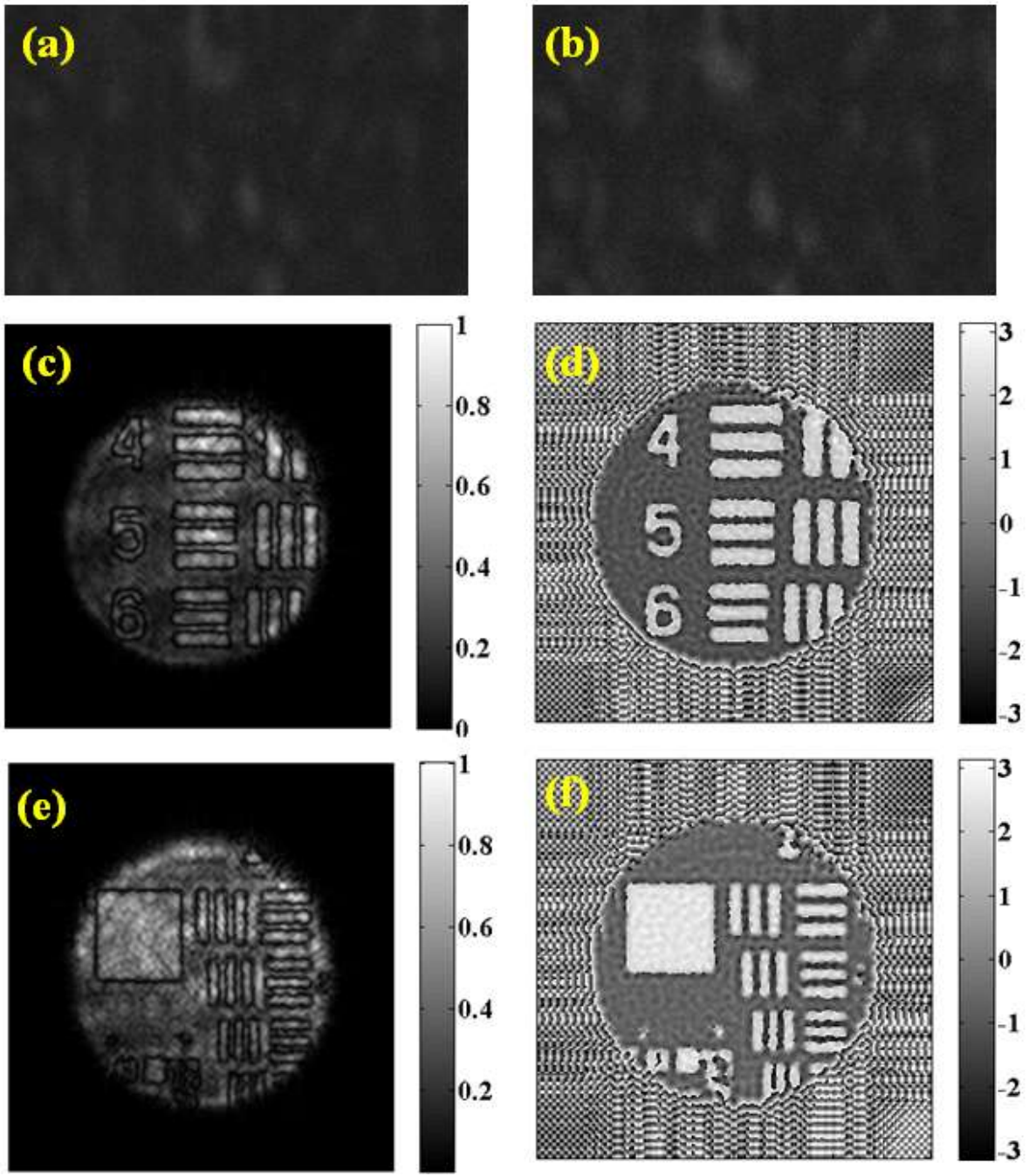

Figure5: Experimental results (a) & (b) random speckle pattern captured by CCD for beam illuminating group 0 and group 1 area of USAF 1951 resolution chart respectively (c) & (e) recovered amplitude distribution of group 0 and group 1 area of USAF 1951 resolution chart (d) & (f) recovered phase distribution of group 0 and group 1 area of USAF 1951 resolution chart

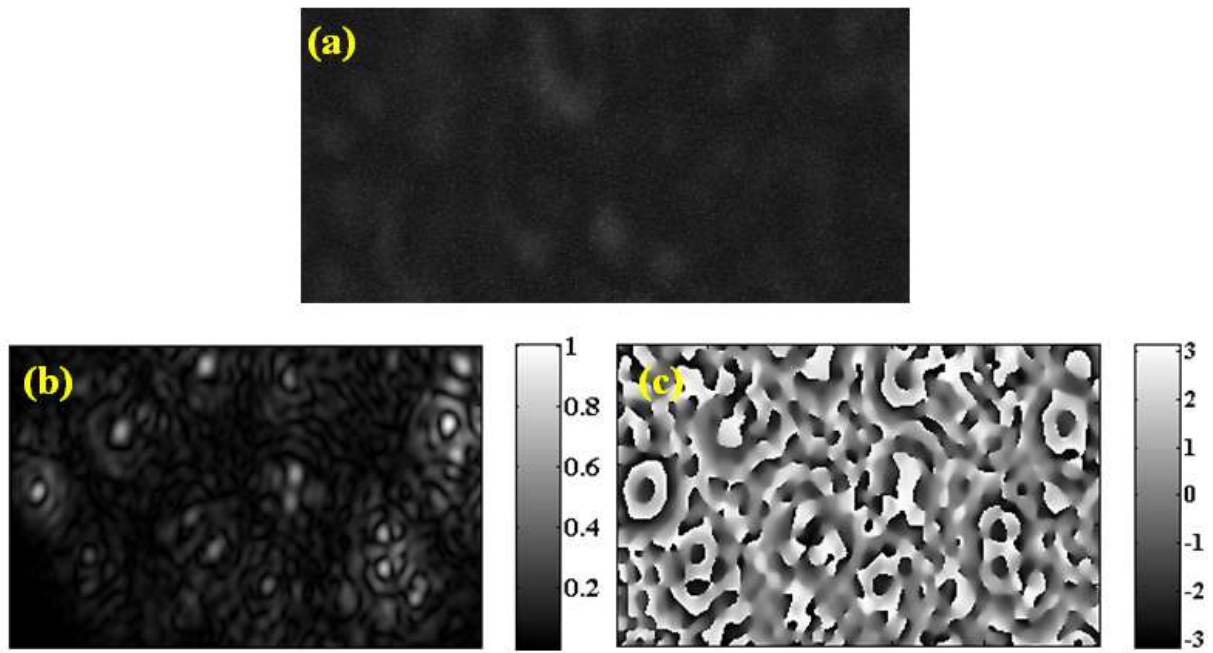

Figure6: Experimental results (a) random speckle pattern recorded by the CCD for a cluster of polystyrene beads immersed in oil and sandwiched between two cover slips as a sample behind the scattering medium